\documentclass[11pt]{article}
\usepackage{graphicx,amsmath,amssymb,bm}

\def\xbr{{\scriptscriptstyle (x)}}
\def\qbr{{\scriptscriptstyle (q)}}
\def\bp {{ \mathbf{p} }} 
\def\bq {{ \mathbf{q} }} 
\def\bx {{ \mathbf{x} }} 
\def\bK {{ \mathbf{K} }} 
\newcommand{\address}[1]{\textit{#1}}
\newcommand{\keywords}[1]{Keywords: #1}

\begin{document}

\begin{center}
  \textbf{\large Determining source cumulants in femtoscopy with
    Gram-Charlier and Edgeworth series}\\[12pt] %
  H.C.\ Eggers, M.B.\ de Kock\\ %
  \address{Department of Physics, University of Stellenbosch,\\
    ZA--7600 Stellenbosch, South Africa}
  \\[12pt]
  \textbf{J.\ Schmiegel}\\ %
    \address{Thiele Centre for Applied Mathematics in Natural
      Sciences, \\
      Department of Mathematics, {\AA}arhus University,\\
      DK–-8000 {\AA}arhus, Denmark} %
\end{center}

\begin{abstract}
  Lowest-order cumulants provide important information on the shape of
  the emission source in femtoscopy.  For the simple case of
  noninteracting identical particles, we show how the fourth-order
  source cumulant can be determined from measured cumulants in
  momentum space. The textbook Gram-Charlier series is found to be
  highly inaccurate, while the related Edgeworth series provides
  increasingly accurate estimates. Ordering of terms compatible with
  the Central Limit Theorem appears to play a crucial role even for
  nongaussian distributions.

  \keywords{femtoscopy, Edgeworth series, correlations, interferometry}
\end{abstract}

PACS Nos.: 13.85.Hd, 13.87.Fh, 13.85.-t, 25.75.Gz

\section{Introduction}

The large experimental statistics which are now available permit
femtoscopic correlations of identical particles (see
e.g. Ref.\cite{Lis05b}) as a function of the full three-dimensional
momentum difference $\bq = \bp_1 - \bp_2$ and often also of the
average pair momentum $\bK = (\bp_1 + \bp_2)/2$. Increasing attention
has therefore been paid to the detailed description in these
higher-dimensional spaces of the second-order correlation function
\begin{eqnarray}
\label{inb} 
1+ R(\bq,\bK) 
&=&  C(\bq,\bK)
\ =\ \frac{\rho(\bq,\bK)}{\rho^{\rm ref}(\bq,\bK)},
\end{eqnarray}
with $\rho$ the density of like-sign pairs in sibling events and
$\rho^{\rm ref}$ the reference pair density usually determined by a
combination of event mixing and Monte Carlo simulation.  After
removing irrelevant correlations,\cite{STAR10a} the correlation
function can yield information on the spacetime statistical properties
of particle emission as embodied in the source function $S(\bx,\bK)$.
The source function is obtained from the two-particle emission
function, the density of particle emission points with near-equal
momenta, by projection from the two four-coordinates onto the relative
three-coordinate $\bx$ (as measured in the pair rest frame of emission
point densities)\cite{Pra84a,Akk02a,Led07a,Dan07a}.  The momentum- and
coordinate-space correlations of noninteracting identical particles in
their centre-of-mass system are related by a Fourier transform
\begin{equation}
  \label{reldist}
  R(\bq,\bK) = \lambda \int d^3x\,S(\bx,\bK)\,e^{i\bq\cdot\bx}\,,
\end{equation}
where $\lambda$ is the correlation strength parameter\footnote{For
  example, $\lambda = 1$ and $(-1/2)$ for noninteracting identical
  spin-0 and unpolarized spin-1/2 particles emitted at independent
  spacetime points respectively. The $\lambda$-parameter may be
  modified by particle impurity or a contribution of the particles
  emitted in some multiparticle quantum states analogous to the
  coherent states in quantum optics. Note that $\lambda$ is eliminated
  by the respective normalisation in Eqs.~(\ref{nmq})--(\ref{nmx}).}.
Suppressing the $\bK$-dependence in a given global system, the
two-particle correlation in both spaces can in this case be written as
a normalised probability density function (pdf) $f(\bq)$ in $q$-space
and a pdf $g(\bx)$ in $x$-space, related by
\begin{eqnarray}
  \label{nmq}
  f(\bq) &=& \frac{R(\bq)}{\int d^3q\, R(\bq)} 
  \ = \ f(\mathbf{0}) \int d^3x\,e^{i\bq\cdot\bx}g(\bx),\\
  \label{nmx}
  g(\bx) &=&
  \frac{S(\bx)}{\int d^3x\,S(\bx)}
  \ =\ 
  \int \frac{d^3q\,e^{-i\bq\cdot\bx}\,f(\bq)}
  {(2\pi)^3 f(\mathbf{0})}
  \ =\
  \frac{S(\bx)}{(2\pi)^3\,S(\mathbf{0})\,f(\mathbf{0})}
       \,.
\end{eqnarray}
A gaussian $f(\bq)$ immediately yields a gaussian $g(\bx)$ in any
dimension. Experimental data, however, is often nongaussian, sometimes
strongly so. This raises two problems: first, to systematically
describe the nongaussian shape of $R$ (or $f$) in momentum space, and
second, to determine parameters of $S$ (or $g$) in coordinate space,
given only the kernel transform and measurements in $q$-space.

Approaches towards systematic description of nongaussian shapes in
$q$-space can be found in e.g.\ Refs.~\cite{Heg93a,Cso00a,Lis05a},
while the source function $S$ is reconstructed by means of
higher-order coefficients in $x$-space using imaging techniques
\cite{Bro97a,Bro98a,Bro05a} and cartesian harmonics
\cite{STAR10a,Dan05a}.

In this paper, we wish to address the second problem of a systematic
description of $g(\bx)$ in terms of given measurements in $q$-space,
based on the fundamental statistical properties of cumulants; the
corresponding approach treating the first problem of measurements in
$q$-space has been treated in part in the literature \cite{Wie96a}
\cite{Egg07a} and will be more fully elaborated elsewhere.

\section{Cumulants in dual spaces}

While fully three-dimensional formulations have been in part set out
in e.g.\ Ref.~\cite{Egg07a}, we shall here work in one dimension using
so that the above expressions become $g(x) = \int
dq\,e^{-iqx} f(q) /2\pi f(0)$ and so on, our purpose being first to test and
improve the convergence properties of series expansions in a simpler
environment.

Given a measured normalised correlation function $f(q)$, its
$q$-moments $\mu_r^\qbr = \int dq\,f(q)\,q^r$ and $q$-cumulants
$\kappa_r^\qbr$ of lowest orders $r = 1,2,3,\ldots$ provide
fundamental information on its properties: the ordinary mean
$\mu_1^\qbr = \kappa_1^\qbr = \int dq\,f(q)\,q$ is a measure of the
\textit{location} of the peak of $f(q)$, while the variance
$\kappa_2^\qbr = \mu_2^\qbr - (\mu_1^\qbr)^2$ measures the dispersion
and $\sigma = (\kappa_2^\qbr)^{1/2}$ the \textit{width} or
\textit{scale} of the pdf, the skewness $\gamma_3^\qbr =
\kappa_3^\qbr/\sigma^3$ measures its \textit{asymmetry} and the
kurtosis $\gamma_4^\qbr = \kappa_4^\qbr/\sigma^4$ is a first
description of the pdf tail's \textit{decay rate}. Higher-order
``generalised kurtoses'' $\gamma_r^\qbr = \kappa_r^\qbr/\sigma^r$
would provide successively more detail.  Kurtoses $\gamma_r^\qbr$ can
also be generally viewed as cumulants of the pdf $f(q')$ of the
standardised variable $q' = (q-\mu_1^\qbr)/\sigma$.

Equivalent relations hold in coordinate space between $x$-moments,
$x$-cumulants and $g(x)$, e.g.\ $\mu_r^\xbr = \int dx\, g(x) \, x^r$,
$\kappa_2^\xbr = \mu_2^\xbr - (\mu_1^\xbr)^2$ and so on.

$q$-moments are derivatives of the generating function $\Phi(x) = 2\pi
f(0) g(-x) = \int dq\,e^{iqx} f(q)$,
\begin{equation}
  \label{mpq}
  \mu_r^\qbr 
  = (-i)^r D_x^r \Phi(x)\bigr|_{x=0}\,,
\end{equation}
writing $D_x^r = (d/dx)^r$ for short, while the related derivation of
$q$-cumulants from
\begin{equation}
  \label{kpq}
  \kappa_r^\qbr 
  = (-i)^r D_x^r  \ln \Phi(x)\bigr|_{x=0}
\end{equation}
fixes relations between moments and cumulants to all orders.
For identical particles, both $C(\bq) = C(-\bq)$ and $g(\bx)$ are
symmetric, so that moments and cumulants of odd order vanish and the
even-order relations in both $q$-space and $x$-space become
\begin{eqnarray}
  \label{kpt}
  \kappa_2 &=& \mu_2, \\
  \label{kpf}
  \kappa_4 &=& \mu_4 - 3 \mu_2^2, \\
  \label{kps}
  \kappa_6 &=& \mu_6 - 15 \mu_4 \mu_2 + 30 \mu_2^3 \qquad \text{etc.}
\end{eqnarray}
Cumulants form a natural basis for near-gaussian expansions since a
gaussian pdf is fully determined once $\kappa_1$ and $\kappa_2$ are
known: all its $\kappa_{r\geq 3}$ are identically zero. They also have
important properties such as invariance under translation and a null
result for uncorrelated variables.

While for purely gaussian sources, the second-order cumulants are
related by $\kappa_2^\xbr = 1/\kappa_2^\qbr$ and all higher-order
cumulants are identically zero, neither of these statements is true in
general. We will therefore consider both the modification of
$\kappa_2^\xbr$ resulting from nonzero $\gamma_r^\qbr$ as well as the
$x$-kurtosis $\gamma_4^\xbr = \kappa_4^\xbr/(\kappa_2^\xbr)^2$.  Since
$x$-moments are found from the generating function
\begin{equation}
  \label{phiq}
  \Phi(q) = f(q) \,/\, f(0)
\end{equation}
through
\begin{equation}
  \label{mpx}
  \mu_r^\xbr 
  = (-i)^r D_q^r \Phi(q)\bigr|_{q=0}\,,
\end{equation}
we can through Eqs.~(\ref{kpt})--(\ref{kps}) obtain $x$-cumulants as
combinations of measured $q$-moments.

\section{Gram-Charlier expansions}

\subsection{Expressing $\gamma_r^\xbr$ in terms of $\gamma_r^\qbr$}

While experimental measurement of derivatives of $\Phi(q)$ is of
course impossible, the above expressions can nevertheless be evaluated
since Gram-Charlier and Edgeworth series expansions also probe regions
of nonzero $q$. Both expansions start with choosing a reference pdf
$f_0(q)$ which, given the close relation between cumulants and
gaussians, is almost invariably chosen by textbooks
\cite{Ken87a,McC87a} to be a gaussian
\begin{equation}
  f_0(q) = \frac{e^{-q^2/2\sigma^2}}{\sigma\sqrt{2\pi}}
  \qquad \text{i.e.}\qquad
  f_0(q')= \frac{e^{-q'^2/2}}{\sqrt{2\pi}}
\end{equation}
with the free parameter $\sigma^2$ fixed to the experimentally
measured $\kappa_2^\qbr$. The resulting ``Gauss Gram-Charlier'' (GGC)
series, also known as the ``Gram-Charlier Type A'' series, and the
corresponding Gauss Edgeworth (GEW) series are closely related, being
mere re-orderings of one another, and are therefore commonly
considered to be one and the same. As we will show, however, the GEW
far outperforms the GGC series at any order of the partial sums.

As shown elsewhere,\cite{Egg06b} the GGC series results from
expanding the generating function for the nongaussian $f(q')$ in
powers of $x'$
\begin{equation}
  \label{ggcphi}
  \Phi(x') = e^{-x'^2/2}\,
  \exp\biggl[\sum_{j=3}^\infty  \zeta_j\, (i x')^j \biggr]
  = e^{-x'^2/2} \sum_{m=0}^\infty \frac{c_m(\bm{\zeta})}{m!} \,(ix')^m
\end{equation}
where each $c_m(\bm{\zeta})$ is a polynomial in the set of
$q$-kurtoses $\bm{\zeta} = \{\zeta_r = \gamma_r^\qbr/r!\}_{r=4}^m$.
Taking the inverse Fourier transform of $\Phi(x')$ term by term, one
obtains an expansion in terms of Chebychev-Hermite polynomials
$H_r(q')$,
\begin{eqnarray}
  \label{xqs}
  f(q') &=& f_0(q') \biggl[ 1 
  + \sum_{j=2}^\infty \frac{c_{2j}(\bm{\zeta})}{(2j)!} H_{2j}(q')
  \biggr], \\
  H_r(q') &=& f_0^{-1}(q')\;(-D_{q'})^r f_0(q')\,,
\end{eqnarray}
with lowest-order terms (writing $H_r(q') = H_r$ for short)
\begin{eqnarray}
  \label{xqqd}
  f(q') &=&  f_0(q') \biggl[
  1 + \zeta_4H_4  + \zeta_6H_6
  + \bigl(\zeta_8+ \tfrac{1}{2}\zeta_4^2\bigr) H_8
  + \bigl( \zeta_{10}+ \zeta_6\,\zeta_4\bigr) H_{10} 
  \nonumber\\
  &&\qquad\qquad 
  +\ \bigl( \zeta_{12} + \zeta_8\,\zeta_4
         + \tfrac{1}{2}\zeta_6^2 + \tfrac{1}{6}\zeta_4^3
    \bigr) H_{12}
  \ +\ \ldots \biggr].
\end{eqnarray}
Using $(-D_{q'})^r f_0(q') H_{2j}(q') = f_0(q') H_{2j+r}(q')$, the
$r$-th derivative of the $x$-moment generating function is, for even
$r$,
\begin{equation*}
  \label{xqc}
  \Phi^{(r)}(q')
  = e^{-q^{\prime 2}/2}
  \left[H_r(q') 
    + \zeta_4 H_{4+r}(q')
    + \zeta_6 H_{6+r}(q') 
    + \ldots
  \right],
\end{equation*}
from which the $x$-cumulants follow as ratios of generating functions
at $q'=0$ in terms of generalised $q$-kurtoses $\gamma_r =
\kappa_r^\qbr/\sigma^r$ and using $H_{2r}(0) = (-1)^r (2r-1)!!$
\begin{eqnarray}
  \label{ktx}
  \kappa_2^\xbr   
  &=& \frac{(-i)^2}{\kappa_2^\qbr} \frac{\Phi^{(2)}(q')}{\Phi^{(0)}(q')}
  \biggr|_{q'=0}
  = \frac{1}{\kappa_2^\qbr}
  \left[\frac{1 
      + \tfrac{5}{8}\gamma_4 
      - \tfrac{7}{48}\gamma_6 
    + \tfrac{3}{128} (\gamma_8 + 35\gamma_4^2) + \ldots}
       {1 + \tfrac{1}{8}\gamma_4 - \tfrac{1}{48}\gamma_6 +
    \tfrac{1}{384} (\gamma_8 + 35\gamma_4^2) + \ldots}    
  \right]
\end{eqnarray}
while the $x$-kurtosis in fourth order is
\begin{eqnarray}
  \label{kfx}
  \gamma_4^\xbr
  &=& \frac{\Phi^{(4)}\,\Phi^{(0)}-3\Phi^{(2)}\,\Phi^{(2)}}
  {\Phi^{(2)}\,\Phi^{(2)}}
  \biggr|_{q'=0}
  =
  \left[
  \frac{
    \gamma_4 
    - \tfrac{1}{2}\gamma_6
    + \tfrac{1}{8}\gamma_8
    + \tfrac{15}{4}\gamma_4^2 
    + \ldots
  }
  {
    1 
    + \tfrac{5}{4}\gamma_4 
    - \tfrac{7}{24}\gamma_6 
    + \tfrac{3}{64}\gamma_8 
    + \tfrac{65}{32}\gamma_4^2 
    + \ldots
  }
  \right]
\end{eqnarray}
with a similar expression for $\kappa_4^\xbr$.

Note firstly that $\gamma_4^\xbr$ depends only on $\gamma_r^\qbr$ but
not directly on $\sigma^2=\kappa_2^\qbr$; this is true also for
higher-order $\gamma_r^\xbr$. Secondly, the above relations reduce to
the gaussian relation $\kappa_2^\xbr = 1/\kappa_2^\qbr$ and
$\gamma_4^\xbr = 0$ if and when the measured correlation function is
gaussian since as mentioned all $\gamma_{r\geq 3}^\qbr$ are then
identically zero. In general, however, the ``radius''
$[2\kappa_2^\xbr]^{1/2}$ of the source distribution is a function also
of higher-order $q$-cumulants, with both increasing orders
$\gamma_r^\qbr$ and increasing powers of lower-order $\gamma_r^\qbr$
entering the expansions.

Given the symmetry between $x$ and $q$, the corresponding expansions
for $\kappa_2^\qbr$ and $\gamma_r^\qbr$ in terms of $\kappa_2^\xbr$
and $\gamma_r^\xbr$ would have the same form as the above, apart from
some changes in sign. Any measured $\kappa_2^\qbr$ is therefore itself
the result of contributions from higher-order cumulants of $g(x)$ or,
in physics terms, the nongaussian shape of the emission region.

\subsection{Truncation and the GGC disaster}

Statistical errors on $q$-cumulants rise with increasing order so that
only those lower-order ones accessible to available experimental
statistics can be included. Series expansions such as (\ref{xqs}) are
known to be asymptotic, so that the question arises: how accurately
can a series truncated at some maximum order $\gamma_{r{\rm max}}$
and/or a maximum power $\gamma_r^{k{\rm max}}$ estimate the
$\gamma_r^\xbr$?

To quantify this issue, we make use of the Normal Inverse Gaussian
(NIG) probability density \cite{Bar97a} as a solvable toy model for
$f(q')$ which yields exact expressions for both coordinate- and
momentum-space cumulants.  While the NIG has four parameters $\alpha$,
$\beta$, $\mu$ and $\delta$, in the present symmetric case
$\beta=\mu=0$, so that we need only the two-parameter Symmetric Normal
Inverse Gaussian (SNIG),
\begin{equation}
  \label{sna}
  f(q\,|\, \alpha,\delta)
  = \frac{\alpha\delta\,e^{\alpha\delta}\,K_1(\alpha\sqrt{\delta^2+q^2})
    } {\pi \sqrt{\delta^2+q^2}}\,,
\end{equation}
where $K_1$ is the modified Bessel function.  The SNIG reverts to a
gaussian in the limit $\alpha\to\infty$ and has $q$-moment generating
function $\Phi(x\,|\,\alpha,\delta) = \exp[\delta\alpha -
\delta\sqrt{\alpha^2 + x^2}]$. Experimentally measured $\kappa_2^\qbr$
and $\gamma_4^\qbr$ would fix the parameters: writing
$\sigma=[\kappa_2^\qbr]^{1/2}$ and $g = \gamma_4^\qbr$ for short,
$\alpha = [3/g\sigma^2]^{1/2}$ and $\delta = [3\sigma^2/g]^{1/2}$, so
that higher-order cumulants and kurtoses can be expressed in terms of
measured quantities $\sigma$ and $g$ as
\begin{eqnarray}
  \label{snb}
  \kappa_{r,\scriptscriptstyle{\rm SNIG}}^\qbr 
  &=& (r-1)!!\,(r-3)!!\;\sigma^r
  \,[\tfrac{1}{3}\,g]^{\tfrac{r}{2}-1}\,,
  \\
  \label{snc}
  \gamma_{r,\scriptscriptstyle{\rm SNIG}}^\qbr 
  &=& (r-1)!!\,(r-3)!!\; [\tfrac{1}{3}\,g]^{\tfrac{r}{2}-1}.
\end{eqnarray}
Using the SNIG pdf as $x$-moment generating function in the form
(\ref{phiq})
\begin{equation*}
  \Phi(q\,|\,\alpha,\delta) 
  = \frac{K_1(\alpha\sqrt{q^2+\delta^2})}{K_1(\alpha\delta)}  \cdot
  \frac{\delta}{\sqrt{q^2+\delta^2}}
\end{equation*}
we obtain exact expressions for $x$-cumulants via (\ref{mpx}) and the
moment-cumulant relations. Omitting the argument of the Bessel
functions, which is $\alpha\delta = 3 / g$ in every case, these
``exact'' $x$-cumulants are
\begin{eqnarray}
  \label{snd}
  \kappa_{2,\scriptscriptstyle{\rm SNIG}}^\xbr
  &=& \frac{1}{\kappa_2^\qbr}\cdot\frac{K_2}{K_1}\,,    \\
  \label{sne}
  \kappa_{4,\scriptscriptstyle{\rm SNIG}}^\xbr 
  &=& \frac{1}{\kappa_2^{\qbr\,2}}\cdot
  \frac{3 K_3 K_1 -3 K_2^2}{K_1^2}\,, \\
  \label{snf}
  \gamma_{4,\scriptscriptstyle{\rm SNIG}}^\xbr 
  &=& \frac{3 K_3 K_1 -3 K_2^2}{K_2^2}\,.
\end{eqnarray}
With these exact $x$-cumulants as reference, we test the accuracy of
various truncations of Eqs.~(\ref{ktx})--(\ref{kfx}) as a
function of the Gram-Charlier order $m=2j$ of Eq.~(\ref{xqs}).

The results are disastrous. In Fig.~1, we show respectively the
percentage deviation of GGC expansions (\ref{ktx}) and (\ref{kfx}),
truncated at $m$th order, from the exact answers (\ref{snd}) and
(\ref{snf}), in the form
$100(\kappa_{2,m}^\xbr/\kappa_{2,\scriptscriptstyle{\rm
    SNIG}}^\xbr-1)$ and
$100(\gamma_{4,m}^\xbr/\gamma_{4,\scriptscriptstyle{\rm
    SNIG}}^\xbr-1)$. At $\gamma_4^\qbr=0$, of course, all series
reduce to a gaussian and all approximations become exact. Even small
values of $\gamma_4^\qbr$ lead to large deviations, however, and the
size of the deviations increases with order $m$. GGC series fail
completely to approximate the exact $x$-cumulants.

\begin{figure}[htb]
\begin{center}
\hspace*{2pt}
\includegraphics[width=75mm]{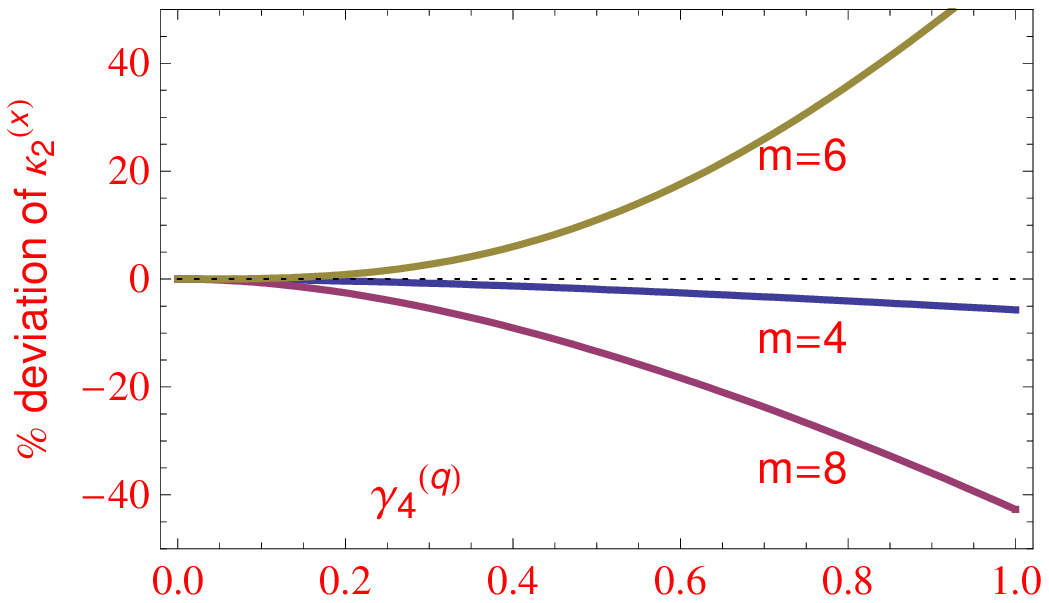}\\
\includegraphics[width=76mm]{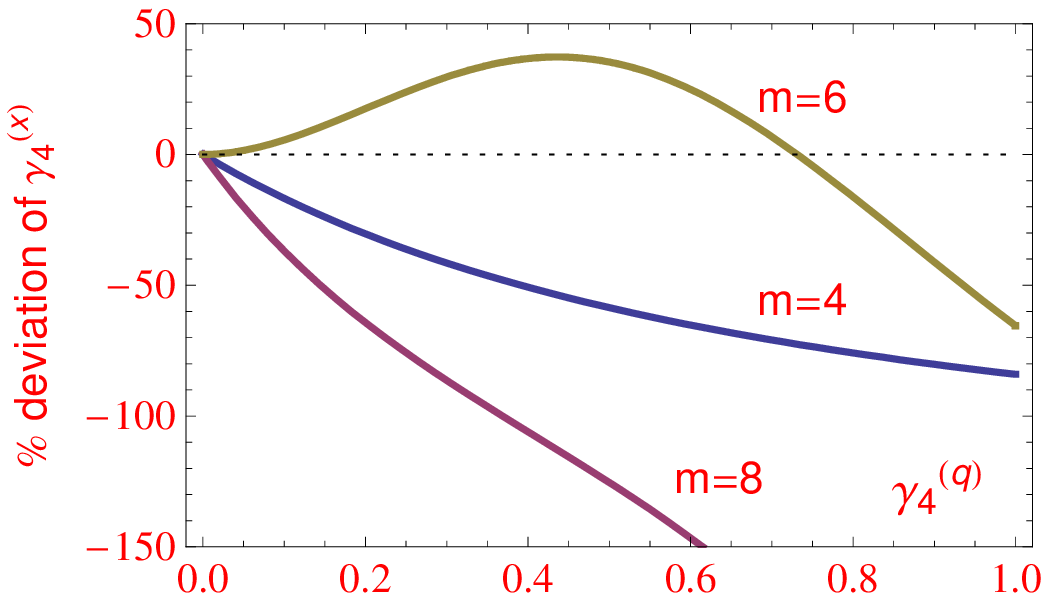}\\
\caption{Gauss Gram-Charlier series fail badly: Percentage deviations
  of Gram-Charlier approximations of $\kappa_2^\xbr$ found by
  Eq.~(\ref{ktx}) vs.\ (\ref{snd}) and of $\gamma_4^\xbr$ found by
  (\ref{kfx}) vs.\ (\ref{snf}), as a function of the measured
  $q$-kurtosis $\gamma_4^\qbr$ for various Gram-Charlier orders $m$. }
\end{center}
\end{figure}

\section{Edgeworth series}

\subsection{Derivation and properties}

In his 1946 treatise on statistics, Cram\'er\cite{Cra46a} derived the
Gauss-Edgeworth (GEW) series by considering the random variable ($q'$
in our case) to be a convolution of $n$ identically-distributed
independent (iid) random variables $q_i$ each with pdf $f_1(q_i)$, a
corresponding generating function $\Phi_1(x_i)$ and second-order
cumulant $\kappa_2^\qbr(n{=}1) = \sigma_1^2$, in terms of which the
generating function for $x'$, the dual to $q' = \sum_i
q_i/(\sigma_1\sqrt{n})$, is
\begin{equation}
  \label{xndv}
  \Phi(x') 
  = \prod_{i=1}^n \int dq_i\,f_1(q_i)\,
    \exp\!\left[\frac{iq_ix'}{\sigma_1\sqrt{n}}\right]
  = \left[\Phi_1\left(\frac{x'}{\sigma_1 \sqrt{n}}\right)\right]^n\,.
\end{equation}
For such convolutions, cumulants of $f(q')$ are related to cumulants
of $f(q_i)$ by $\kappa_j^\qbr(n) = n\,\kappa_j^\qbr(1)$, so that
\begin{equation}
  \label{zetaconv}
  \zeta_j(n) = \frac{\zeta_j(1)}{n^{\tfrac{j}{2}-1}}\,.
\end{equation}
and hence $\Phi(x')$ depends on $\ell = n^{-1/2}$ through
\begin{equation}
  \Phi(x') 
  = e^{-x'^2/2}
  \exp\biggl[\, \sum_{j=3}^{\infty} (ix')^j\,\zeta_j(n)
    \biggr]
  \ =\ e^{-x'^2/2}
  \exp\biggl[\, \sum_{j=3}^{\infty} (ix')^j\,\zeta_j(1)\,\ell^{j-2}
    \biggr].
\end{equation}
Expanding the exponential in powers of $\ell$ rather than of $x'$ and
again inverting term by term, we obtain the Gauss-Edgeworth series
\begin{eqnarray}
  \label{ewz}
  f(q') 
  &=& f_0(q') \biggl[
    1
    +\ell^2 \zeta_4 H_4
    +\ell^4 \left(
      \tfrac{1}{2}\zeta_4^2  H_8
      +  \zeta_6 H_6 \right) 
    + \ell^6 \left( 
      \tfrac{1}{6}\zeta_4^3  H_{12} 
      + \zeta_4\zeta_6 H_{10} 
      +  \zeta_8 H_8
    \right)
    \nonumber\\
    &&\quad + \ell^8 \left(
      \tfrac{1}{24}\zeta_4^4   H_{16}
      + \tfrac{1}{2}\zeta_4^2\zeta_6  H_{14}
      + \tfrac{1}{2}\zeta_6^2  H_{12}
      +  \zeta_4 \zeta_8 H_{12}
      +  \zeta_{10} H_{10}
    \right) 
    \ +\ \ldots
    \biggr],
\end{eqnarray}
again writing $H_r = H_r(q')$ for short and with $\zeta_j$ here
understood as $\zeta_j(1)$.  Unlike the equivalent GGC expansion of
Eq.~(\ref{xqqd}), in which the order of the expansion was determined
by the order of $H_m$, a given term of order $\ell^w$ in the GEW
series is a linear combination of Hermite polynomials of different
order.

The relation between Gram-Charlier and Edgeworth ordering is
summarised in Table 1, with terms listed in ascending order for $w$.
A given term $\zeta_{r_1}\zeta_{r_2}\!\cdots$ is characterised by the
set of partition coefficients $r_k = 4,6,8,\ldots$ which are
constrained to the Gram-Charlier and Edgeworth orders by
\begin{eqnarray}
  \sum_k r_k &=& m\,, \\
  \sum_k (r_k - 2) &=& w\,.
\end{eqnarray}
The re-ordering becomes important already for the second-lowest 
order $w=4$.\\

\begin{center}
\renewcommand{\arraystretch}{1.15}
\setlength{\tabcolsep}{4pt}
\begin{tabular}{|c|r|c|c|}
   \hline
   series  &         & Edgeworth        & Gram-Charlier           \\
   term    & $\{r_k\}\quad$ &  order $w$ & order $m$  \\
   \hline\hline
   $\zeta_4$          & $\{4\}$   & 2 & 4   \\ \hline
   $\zeta_6$          & $\{6\}$   & 4 & 6   \\ \hline
   $\zeta_4^2$        & $\{4,4\}$ & 4 & 8   \\ \hline
   $\zeta_8$          & $\{8\}$   & 6 & 8   \\ \hline
   $\zeta_6\zeta_4$  & $\{6,4\}$ & 6 & 10  \\ \hline
   $\zeta_4^3$        & $\{4,4,4\}$ & 6 & 12  \\ \hline
   $\zeta_{10}$        & $\{10\}$ & 8 & 10 \\ \hline
   $\zeta_6^2$         & $\{6,6\}$ & 8 & 12 \\ \hline
   $\zeta_8\zeta_4$   & $\{8,4\}$ & 8 & 12  \\ \hline
   $\zeta_6\zeta_4^2$ & $\{6,4,4\}$ & 8 & 14  \\ \hline
   $\zeta_4^4$         & $\{4,4,4,4\}$ & 8 & 16  \\ \hline
\end{tabular}
\\[6pt]
\noindent {\footnotesize Table 1: Re-ordering of terms between
  Gram-Charlier (GC) and Edgeworth (EW) series}\\[12pt]
\end{center}

\subsection{Test using SNIG}

Edgeworth re-ordering of terms in the derivatives $\Phi^{(r)}(q')$
leads to expressions for the $x$-cumulants as ratios of power
series\footnote{Since $\kappa_4^\xbr$ and $\gamma_4^\xbr$ contain
  products of generating functions, terms of order higher than
  $\ell^w$ are generated. Such terms must of course be omitted in a
  consistent $O(\ell^w)$ calculation.}  in $\ell$. For the SNIG test
case, these series simplify to
\begin{eqnarray}
  \label{ketx}
  \kappa_2^\xbr  
  &=& 
  \frac{1}{\kappa_2^\qbr}
  \left[\frac{1+\frac{5}{8} \gamma_4 \ell^2  +\frac{35}{384} \gamma_4^2 \ell^4
      -\frac{35}{3072} \gamma_4^3 \ell^6
      +\frac{385}{98304} \gamma_4^4 \ell^8    + \ldots}
  {1+\frac{1}{8}\gamma_4\ell^2
    -\frac{5}{384} \gamma_4^2 \ell^4
    +\frac{35}{9216} \gamma_4^3\ell^6
    -\frac{175}{98304}  \gamma_4^4\ell^8   + \ldots}
  \right],
  \\[6pt]
  \label{kefx}
  \gamma_4^\xbr
  &=&\qquad
  \left[
    \frac
    {\gamma_4\ell^2 
      +\frac{5}{4} \gamma_4^2\ell^4
      +\frac{35}{96} \gamma_4^3\ell^6
      -\frac{35}{1152}  \gamma_4^4\ell^8   + \ldots}
    {1+\frac{5}{4} \gamma_4\ell^2
      +\frac{55}{96} \gamma_4^2\ell^4
      +\frac{35}{384} \gamma_4^3\ell^6
      +\frac{35}{18432} \gamma_4^4\ell^8    + \ldots}
  \right]\,,
\end{eqnarray}
with $\gamma_4 \equiv \gamma_4^\qbr(1)$.  The algebraic simplicity of
the above compared to the equivalent GGC relations
(\ref{ktx})--(\ref{kfx}) and the GEW relation (\ref{ewz}) is due to
the fact that SNIG kurtoses obey
\begin{equation}
  \label{snigpower}
  \gamma_j^\qbr = F_j\, [\gamma_4^\qbr]^{\tfrac{j}{2}-1}
\end{equation}
with constants $\{F_4,F_6,F_8,\ldots\} =
\{1,5,\tfrac{175}{3},\ldots\}$ fully determined by the SNIG pdf. While
relation (\ref{snigpower}) is of course fulfilled by all convolutions
through (\ref{zetaconv}), it is true for the SNIG case even without
convolution.

In Fig.~2, we show the percentage deviations of the
Edgeworth-truncated approximations (\ref{ketx})--(\ref{kefx}) from
their respective exact SNIG values as a function of the $q$-kurtosis
$\gamma_4^\qbr$. The improvement in accuracy over the GGC ordering is
dramatic. Unlike the GGC, the GEW series also continues to improve
as higher orders of $w$ are included.

\begin{figure}[htb]
\begin{center}
\includegraphics[width=77mm]{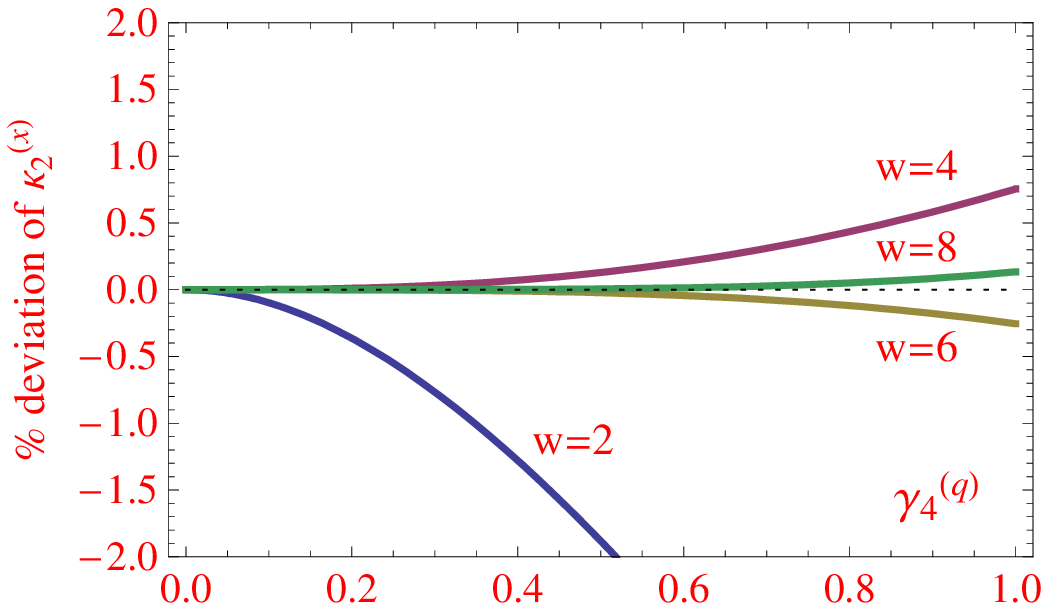}\\
\includegraphics[width=76mm]{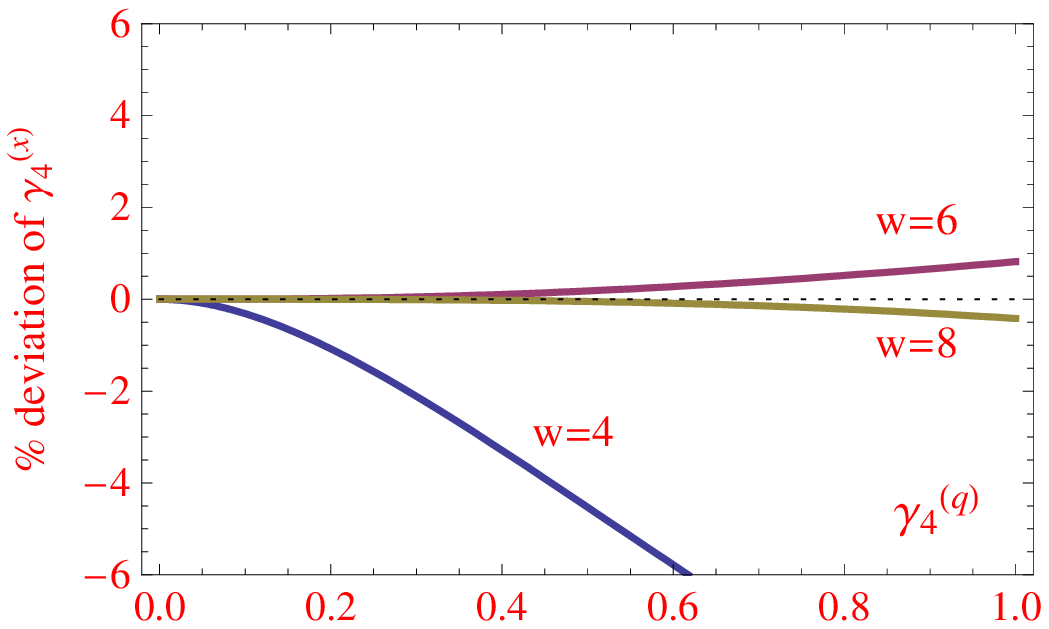}\\
\hspace*{2pt}
\caption{Gauss Edgeworth series: Percentage deviations of GEW
  approximations of $\kappa_2^\xbr$ found by Eq.~(\ref{ketx}) vs.\
  (\ref{snd}) and of $\gamma_4^\xbr$ found by (\ref{kefx}) vs.\
  (\ref{snf}), as a function of the $q$-kurtosis
  $\gamma_4^\qbr$ for various Edgeworth orders $w$. The dashed line at
  zero marks perfect agreement between the truncated expansion and
  the exact SNIG answer.}
\end{center}
\end{figure}

\subsection{$n$-divisibility and the Central Limit Theorem}

What structure or principle underlies the strong superiority of GEW
over GGC ordering? Clearly, the expansion parameter $n = \ell^{-2}$
must be playing a crucial role.  Eq.~(\ref{xndv}) characterises the
generating function $\Phi(x')$ and therefore the pdf $f(q')$ as
``$n$-divisible'' or more prosaically as precisely that convolution
$f(q') = \delta[q'-\sum_i q_i/\sigma_1\sqrt{n}\,]\,\prod_i f(q_i)$ which
led to Eq.~(\ref{xndv}). The GEW method might therefore be expected to
work well if $n$-divisibility could be established for a given
experimental data set. However, it is usually not possible to directly
establish whether the $f(q')$ of an experimental data set is
$n$-divisible, and there is no obvious physics reason to believe that
a momentum difference $q'$ between two particles is the result of an
underlying summation.

The reason for the success of GEW relies not on physics assumptions,
but on a better description of nongaussian systems. In statistics
terms, any deviation from the gaussian is captured by the GEW series
in the \textit{rate of approach} of higher-order cumulants to zero
with increasing $n$. Indeed, many proofs of the Central Limit Theorem
rely on the fact that $\kappa_j(n) = n\,\kappa_j(1)$ immediately
yields the fixed rate of convergence for any generalised kurtosis
as shown in Eq.~(\ref{zetaconv}).

The success of GEW ordering is therefore based on the fact that all
contributing terms in a given order of $\ell = n^{-1/2}$ have the same
rate of convergence to the gaussian limit. Furthermore, due to the
alternating sign of $H_{2r}(0) = (-1)^r (2r-1)!!$, the sum of
contributions within a given $O(\ell^w)$ term tends to be
substantially smaller than the individual contributions; for example
the $w=4$ term for the SNIG test case is made up of
$\tfrac{1}{2}\zeta_4^2 H_8(0) = 0.091 \gamma_4^2$ and $\zeta_6F_6
H_6(0) = -0.104 \gamma_4^2$, adding up to $-0.013 \gamma_4^2$.

It is not necessary to know the value of $n$ to make use of GEW
ordering: Using (\ref{zetaconv}), we can merely re-absorb the $\ell^w$
in (\ref{ewz}) to write the GEW expansion in $n$-independent notation
(now with $\zeta_j \equiv \zeta_j(n)$, the measured kurtosis)
\begin{eqnarray}
  \label{ewzb}
  f(q') 
  &=& f_0(q') \biggl[
    1
    +\bigl( \zeta_4 H_4\bigr)
    +\left(
      \tfrac{1}{2} \zeta_4^2  H_8
      +  \zeta_6 H_6 \right) 
    +\left( 
      \tfrac{1}{6} \zeta_4^3   H_{12} 
      + \zeta_4\zeta_6 H_{10} 
      +  \zeta_8 H_8
    \right)
    \nonumber\\
    &&\quad  + \left(
      \tfrac{1}{24} \zeta_4^4   H_{16}
      + \tfrac{1}{2} \zeta_4^2\zeta_6  H_{14}
      + \tfrac{1}{2} \zeta_6^2   H_{12}
      +  \zeta_4 \zeta_8 H_{12}
      +  \zeta_{10} H_{10}
    \right) 
    \ +\ \ldots
    \biggr],
\end{eqnarray}
but keeping in mind that each expression within a round bracket
represents a certain rate of convergence and must be included or
excluded as a whole.

It is not even necessary to require $n$-divisibility as such: For the
GEW ordering to be effective we require only that $f(q)$ is reasonably
close to a gaussian, where ``reasonable'' is typically quantified by
the errors shown in Fig.~2.  The derivation also does not rely on a
particular form of $f_1(q_i)$ other than requiring existence of its
cumulants.

\section{Conclusions}

We have calculated expressions for $x$-cumulants of second and fourth
order for the emission region in terms of measured $q$-cumulants. On
using a nongaussian test function to quantify accuracy of expansions,
we have shown that the textbook Gram-Charlier series is unsuitable at
any level of approximation. By contrast, the Gauss-Edgeworth
expansion, which orders terms based on the rate of approach to a
gaussian, does give results which become increasingly accurate as more
terms are added. The GEW series is robust in all respects; for example
it does not require $n$-divisibility as such but only that $f(q)$ be
close enough to a gaussian in the sense that the measured
$q$-cumulants entering truncated GEW expansion should have reasonable
error bars.

The present one-dimensional calculation clearly cannot be applied
immediately to experimental data, but is meant to show that, even on
the fundamental level of expansions, there are major questions which
must be addressed first. In sorting out the fundamental issue of
re-ordering, the present results represent an important step towards a
consistent framework for shape description.

Application to experimental data will require generalisation to three
dimensions using the existing 3D machinery of Refs
\cite{Egg07a,Egg06b}. Furthermore, sampling fluctuations of
experimental cumulants will have to be taken into account. In this
connection, we also note that the GEW ordering has the additional
advantage of placing terms with higher powers of $\gamma_4^\qbr$ into
lower orders of $w$, making it unnecessary to measure higher-order
kurtoses. Based on slightly different arguments, Cram\'er
\cite{Cra46a} also concluded that GEW was superior to GGC ordering;
this has also been verified GGC vs GEW comparisons of the nongaussian
pdf itself \cite{Bli98a,DeK09a}. The Gram-Charlier series can
therefore be considered to be inferior to the Edgeworth equivalent in
all respects.

Note that it is not necessary to measure the correlation function
$f(q)$ at $q=0$, despite the fact that $x$-cumulants rely formally on
the generating function (\ref{mpx}) at zero. The $q$-cumulants
themselves are functions of $f(q)$ over the whole range of $q$, while
the correlation strength parameter $\lambda$ cancels in the
normalisation (\ref{nmq}).

Our final comment pertains to the usual practice of obtaining
information on the correlation function through fits of nongaussian
parametrisations. Fits rely on an \textit{a priori} choice of
parametrisation, guided only by the minimisation of $\chi^2$, and
suffer from increasing ambiguity in higher dimensions. By relying on
direct measurement of coefficients, the present method and those of
Refs \cite{Bro98a,Bro05a} etc leave less room for arbitrary choices
and put the uncertainty where it belongs: in the sampling fluctuations
of measured experimental quantities.

\textbf{Acknowledgements:} This work was supported in part by the
National Research Foundation of South Africa.


\end{document}